\documentclass[twocolumn,english,showpacs,floatfix,amsmath,amssymb,prb]{revtex4-1}
\usepackage[T1]{fontenc}
\usepackage[latin9]{inputenc}
\usepackage{babel}

\usepackage{graphicx}
\usepackage{amssymb}
\usepackage{esint}

\usepackage{bm}
\usepackage{dsfont}
\usepackage{ulem}
\usepackage[usenames]{color}
\usepackage{soul}
\setstcolor{red}
\sethlcolor{yellow}



\begin{document}

\title{Topological superconducting phases in disordered quantum wires with strong spin-orbit coupling}

\author{Piet W.\ Brouwer} 
\affiliation{\mbox{Dahlem Center for Complex Quantum Systems and Fachbereich Physik, Freie Universit\"at Berlin, 14195 Berlin, Germany}}

\author{Mathias Duckheim} 
\affiliation{\mbox{Dahlem Center for Complex Quantum Systems and Fachbereich Physik, Freie Universit\"at Berlin, 14195 Berlin, Germany}}

\author{Alessandro Romito}
\affiliation{\mbox{Dahlem Center for Complex Quantum Systems and Fachbereich Physik, Freie Universit\"at Berlin, 14195 Berlin, Germany}}

\author{Felix von Oppen}
\affiliation{\mbox{Dahlem Center for Complex Quantum Systems and Fachbereich Physik, Freie Universit\"at Berlin, 14195 Berlin, Germany}}

\date{\today}
\begin{abstract}
Zeeman fields can drive semiconductor quantum wires with strong spin-orbit coupling and in proximity to $s$-wave superconductors into a topological phase which supports end Majorana fermions and offers an attractive platform for realizing topological quantum information processing. Here, we investigate how potential disorder affects the topological phase by a combination of analytical and numerical approaches. Most prominently, we find that the robustness of the topological phase against disorder depends sensitively and non-monotonously on the Zeeman field applied to the wire. 
\end{abstract}
\maketitle

\section{Introduction}

Topological quantum information processing promises to go a long way towards alleviating the problem of environmental decoherence in quantum computers. \cite{kitaev03,freedman98} In this scheme, information is stored and processed by nonlocal qubits based on quasiparticles with nonabelian quantum statistics. \cite{frohlich90,moore91} These nonlocal qubits are topologically protected against local perturbations. Qubit operations are based on quasiparticle exchanges which are themselves topological in nature and thus insensitive to small variations in the operation. 

The simplest type of quasiparticles obeying nonabelian quantum statistics are zero-energy Majorana fermions. \cite{review,wilczek} Such Majorana bound states are carried by vortices of weak-pairing superconductors with spinless $p$-wave pairing symmetry, a phase which is widely expected to be realized by the $\nu=5/2$ fractional quantum Hall state. \cite{read00} Starting with the seminal suggestion of Fu and Kane \cite{fu08} that zero-energy Majorana fermions can in principle be engineered to exist in hybrid structures of conventional $s$-wave superconductors and topological insulators, various superconductor hybrids have been predicted to support Majorana fermions. These hybrids involve, e.g., standard two-dimensional electron systems with strong spin-orbit coupling, \cite{sau10} metallic surface states, \cite{potter10} semiconductor quantum wires with strong spin-orbit coupling, \cite{lutchyn10,oreg10} or half-metallic wires and films. \cite{lee09,duckheim11,chung11}

Realizations based on quantum wires are perhaps particularly promising for realizing topological quantum information processing since they allow for relatively detailed scenarios \cite{alicea11} of how to manipulate (i.e., create, transport, or fuse) Majorana fermions. At the same time, disorder is known to often have drastic consequences for the electronic properties of one-dimensional electron systems \cite{beenakker97} and for superconducting pairing correlations. \cite{degennes} This motivates us to address the important question of how robust these proposed realizations of Majorana fermions are against potential disorder. 

The effects of disorder on topological phases  which support Majorana fermions were studied in relation to fermion zero modes inside the vortex core in p-wave superconductors in the presence of impurities by Volovik~\cite{Volovik1999} and, in the one-dimensional case, by Motrunich {\it et al.}. \cite{motrunich01} Both papers predate the current wave of interest by almost a decade. Motrunich {\it et al.} consider a disordered $p$-wave superconductor in one dimension within a renormalization group approach. Their central finding is that disorder causes a sharp transition to a non-topological phase at a critical disorder strength. More recent work on disorder effects on Majorana fermions focuses on the strong-disorder limit where the wire breaks up into topological and non-topological domains. \cite{flensberg10,shivamoggi10} 

It is the central purpose of the present paper to emphasize that the physics of disorder in proximity-coupled semiconductor wires with strong spin-orbit coupling is considerably richer. It reduces to the model studied by Motrunich {\it et al.} (sometimes refered to as Kitaev's toy model) \cite{kitaev01} only in the limit of a large Zeeman field. In contrast, we find that the semiconductor quantum wires exhibit additional regimes, depending on the strength of the Zeeman coupling, which display characteristic differences in the effectiveness of disorder. Most importantly, our results can give significant guidance to efforts to realize Majorana fermions in the laboratory. 

\begin{figure*}[t]
\includegraphics[width=16cm, keepaspectratio=true]{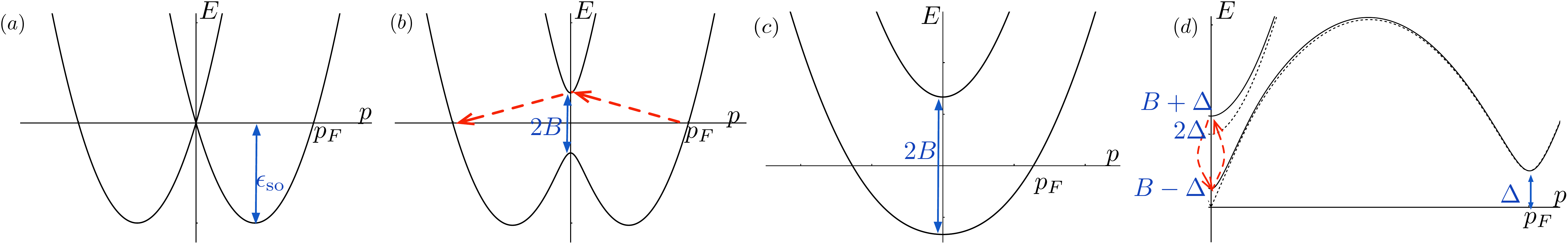} 
\caption{(Color online)
Dispersion relations in the absence of disorder for: (a) the quantum wire in the normal phase and zero Zeeman field; (b) the quantum wire in the normal phase and finite Zeeman field $B\ll \varepsilon_{\rm so}$, including an illustration (red arrows) of a second-order process contributing to the effective disorder potential in the regime of intermediate Zeeman fields; (c) the quantum wire in the normal phase and strong Zeeman field $B\gg \varepsilon_{\rm so}$. (d) Quasiparticle excitation spectrum near (full line) and at (dashed line) the topological phase transition $\Delta=B$. The red (dashed) arrows illustrate a second-order process to the high-energy subspace near $p=0$, which contributes to the random Zeeman field in the low-energy model. 
\label{fig:dispersions}}
\end{figure*}

\section{Semiconductor wires}

We consider a semiconductor quantum wire with strong Rashba spin-orbit coupling $u$ in an external Zeeman field $B$. For definiteness, we take the spin-orbit field and the Zeeman field to point along the $x$- and $z$-directions, respectively. If the wire is in proximity to an $s$-wave superconductor, the corresponding Bogoliubov-de Gennes (BdG) Hamiltonian takes the form \cite{lutchyn10,oreg10}
\begin{equation}
  {\cal H} = \left( \frac{p^2}{2m} + u p \sigma_x + V(x) - \mu \right)\tau_z - B \sigma_z + \Delta \tau_x . 
  \label{BdG}
\end{equation}
Here, $p$ and $x$ denote the momentum and position along the wire and $\Delta$ is the proximity-induced pairing potential, i.e.\ the gap in the absence of $V$, $\mu$, $B$. 
The Pauli matrices $\sigma_i$ ($\tau_i$) operate in spin (particle-hole) space. Disorder is modeled through a Gaussian white noise potential $V(x)$ with zero mean and correlator $\langle V(x)V(x^\prime)\rangle = \gamma\delta(x-x^\prime)$. \cite{gangadharaiah11} The disorder strength $\gamma$ is related to the mean scattering time $\tau_0 = u/\gamma$ at $\mu=B=0$.

Eq.\ (\ref{BdG}) is written in a basis which corresponds to the four-component Nambu operator $\Psi = [\psi_\uparrow, \psi_\downarrow, \psi^\dagger_\downarrow,-\psi^\dagger_\uparrow]$ in terms of the electronic field operator $\psi_\sigma(x)$. In this basis, the time-reversal operator takes the form $T = i \sigma_y K$, where $K$ denotes complex conjugation. The BdG Hamiltonian (\ref{BdG}) obeys the symmetry $\{{\cal H},CT\} = 0$, with $C = - i\tau_y$. Thus, if $|\psi\rangle$ is an eigenspinor of ${\cal H}$ with energy $E$, $CT|\psi\rangle$ is an eigenspinor of energy $(-E)$. Zero-energy Majorana fermions are characterized by spinors satisfying ${\cal CT}|\psi_0\rangle = |\psi_0\rangle$,  Majorana modes by ${\cal CT}|\psi_p\rangle = |\psi_{-p}\rangle$.

\section{Regimes and effective low-energy models}

In the absence of disorder and of the proximity-induced gap, i.e., for $\Delta = V(x) = 0$, the Hamiltonian Eq.\ (\ref{BdG}) has the particle dispersion $E_{p,\pm}^{(0)} = p^2/2m \pm [(up)^2+B^2]^{1/2}$. A topological superconducting phase requires that there are only two Fermi points, cp.\ Figs.\ \ref{fig:dispersions}(a)-(c). In this case, the chirality dependence of the spin orientation induced by the Rashba coupling turns the normal state into a helical liquid. \cite{lutchyn10,oreg10} The superconducting phase for $B<\Delta$ is a conventional, non-topological $s$-wave superconductor. A topological superconducting phase can be realized when the Zeeman field is larger than $\Delta$. For strong spin-orbit coupling $\epsilon_{\rm so} \gg \Delta$ (with the spin-orbit energy $\epsilon_{\rm so}= mu^2/2 $), it is natural to distinguish three regimes:

\subsection{Weak Zeeman fields ($\epsilon_{\rm so} \gg B>\Delta$ with $B-\Delta \ll \Delta$)} 
\label{(i)}
The gap at $p=0$ is given by $b = B-\Delta$ in the presence of both Zeeman field and proximity-induced superconductivity. Thus, as long as $|B-\Delta| \ll \Delta$, the gap at $p=0$ is much smaller than the proximity-induced gap at the Fermi points which is equal to $\Delta$. In this case, the relevant low-energy degrees of freedom for $\mu\simeq 0$ are those with momenta $p$ near $p=0$. \cite{oreg10} (Note that here we take $\Delta$ and $B$ to be positive.)

We can develop a low-energy model by focusing on small $p$ where $p^2/2m$ can be neglected relative to the spin-orbit term and expanding about the critical point $B=\Delta$ at which the $p=0$ gap closes. The low-energy subspace is then spanned by the eigenspinors $\langle x|p,\pm\rangle = (1/2\sqrt{L})[1,\pm 1,1,\mp 1]^T e^{ipx}$ of the BdG Hamiltonian, and the Hamiltonian evaluated in this subspace takes the form
\begin{equation}
    {\cal H} = \left( \begin{array}{cc} up - v & -b + w \\ -b + w & -up + v \end{array} \right)
\label{WeakZeeman}
\end{equation}
Remarkably, all matrix elements of the disorder potential in the low-energy subspace vanish so that to this order, we find that the disorder fields $v = w =0$. Note that the eigenspinors of the low-energy subspace correspond to Majorana modes. 

The leading contributions to the disorder fields $v$ and $w$ are quadratic in the bare disorder potential, emerging from virtual excitations of high-energy states shown in Fig.\ \ref{fig:dispersions}(d). Adding the contributions from the three relevant high-energy subspaces near $p=0$ and $p=\pm p_F$ ($p_F=2mu$), we obtain
\begin{eqnarray}
  v &=&  V \frac{up}{ u^2p^2 + 4\Delta^2} V - \sum_{\pm} V\frac{u(p\pm p_F)/2}{u^2(p\pm p_F)^2+\Delta^2}V,\\
  w &=&  V\frac{2\Delta}{u^2p^2 + 4\Delta^2} V + \sum_{\pm} V \frac{\Delta/2}{u^2(p\pm p_F)^2+\Delta^2}V.
\end{eqnarray}
In the low-energy limit, the random gauge field $v(x)$ vanishes, while $w(x)$ is a Gaussian white noise Zeeman field with $\langle w(x)\rangle = \gamma/u$ and $\langle  w(x) w(x^\prime)\rangle - \langle w(x)\rangle^2 = (\gamma^2/2u\Delta) \delta(x-x^\prime)$. The effective Hamiltonian Eq.\ (\ref{WeakZeeman}) remains a valid approximation for $\gamma\ll\gamma_{\rm max}\sim u \Delta$.

\subsection{Intermediate Zeeman fields ($\varepsilon_{\rm so}\gg B \gg \Delta$)} 
\label{(ii)}
When the Zeeman field is much larger than the proximity-induced gap $\Delta$, but still small compared to the spin-orbit energy $\varepsilon_{\rm so}$, the gap at $p=0$ is of order $B$ and thus much larger than the gap at the Fermi points of order $\Delta$. In this case, the relevant low-energy degrees of freedom for $\mu\simeq 0$ are those near the Fermi points $\pm p_F$. It is important to realize that the spin orientation at the Fermi points is dominated by the spin-orbit coupling so that electrons at the two Fermi points have almost antiparallel spins. Thus, the proximity effect is not attenuated by spin effects, while disorder-induced backscattering is strongly suppressed. 

This can be made explicit by projecting the original Hamiltonian Eq.\ (\ref{BdG}) onto the lower band of the normal-state Hamiltonian of the clean wire. The corresponding electron and hole eigenspinors are $|p,{\rm e}\rangle= [\cos (\alpha/2),-\sin(\alpha/2),0,0]^T e^{ipx}/\sqrt{L}$ and $|p,{\rm h}\rangle = [0,0,-\sin (\alpha/2),\cos(\alpha/2)]^T e^{ipx}/\sqrt{L}$ with $\tan\alpha = up/B$. Evaluating the matrix elements of the full Hamiltonian in this low-energy subspace and linearizing the dispersion about the Fermi points, we find a spinless $p$-wave superconductor 
\begin{equation}
{\cal H} = v_{\rm eff} (p\lambda_z - p_F)\tau_z + \Delta_{\rm eff}\lambda_y \tau_x + V_{\rm eff}(x) \lambda_x \tau_z,
\label{Kitaev} 
\end{equation}
with Fermi momentum $p_F=2mu$, Fermi velocity $v_{\rm eff}=u$, and gap function $\Delta_{\rm eff}=\Delta$. The disorder potential has the correlation function $\langle V_{\rm eff}(x) V_{\rm eff}(x')\rangle = \gamma_{\rm eff} \delta(x-x')$, corresponding to the scattering time $\tau_{\rm eff} = v_{\rm eff}/\gamma_{\rm eff}$. The Pauli matrices $\lambda_i$ operate in the space of left- and right-movers. In Eq.\ (\ref{Kitaev}), we ignore forward scattering by disorder, which is not expected to affect the results. 

Unlike for weak Zeeman fields, the effective disorder potential $V_{\rm eff}(x)$ has a contribution $V_{\rm eff}(x) = (B/4\varepsilon_{\rm so})V(x)$ [such that $\gamma_{\rm eff} = (B/4\varepsilon_{\rm so})^2\gamma$] which is linear in the bare disorder potential. However, this contribution includes a strong  suppression by spin effects and for strong enough disorder $\gamma \gg \gamma_{\rm lin}\sim (Bu)(B/\varepsilon_{\rm so})^2$, the dominant contribution to $V_{\rm eff}$ remains quadratic in the bare disorder potential, involving virtual intermediate states near $p=0$, cp.\ Fig.\ \ref{fig:dispersions}(b). We find that the corresponding magnitude of the effective disorder potential is $\gamma_{\rm eff} = \gamma^2/2Bu$. Note that for intermediate Zeeman fields, the projection to the lower subband remains well defined for $\gamma\ll Bu$.

\subsection{Strong Zeeman fields ($B \gg \varepsilon_{\rm so} \gg \Delta$)} 
\label{(iii)}
The low-energy theory in this limit has been derived previously for the clean case and can be obtained as in the case of intermediate Zeeman fields. \cite{alicea11} The relevant degrees of freedom remain those in the vicinity of the Fermi points [cp.\ Fig.\ \ref{fig:dispersions}(c)], but the spin orientation is now dominated by the Zeeman field and electrons at the two Fermi points have essentially parallel spin. Consequently, the proximity effect is suppressed, while backscattering is controlled directly by the bare disorder potential $V(x)$. Indeed, the projection [valid for $\gamma\ll Bu(B/\epsilon_{\rm so})^{1/2}$] readily yields a spinless $p$-wave superconductor Eq.\ (\ref{Kitaev}). Unlike in the regimes of weak and intermediate Zeeman fields, the detailed parameters differ for the cases of fixed chemical potential $\mu\simeq 0$ and fixed density $n=2mu/\pi$. For fixed $\mu$, one finds $p_F = (2mB)^{1/2}$, $v_{\rm eff} = (2B/m)^{1/2}$, and $\Delta_{\rm eff}=2\Delta(\varepsilon_{\rm so}/B)^{1/2}$. For fixed $n$, we have $p_F=2mu$, $v_{\rm eff}=u$, and $\Delta_{\rm eff} = 4 \Delta(\varepsilon_{\rm so}/B)$. In both cases, $V_{\rm eff}(x)=V(x)$

\section{Phase diagram}

Equation (\ref{Kitaev}) is the continuum version of the lattice model studied by Motrunich {\it et al.}. \cite{motrunich01} It has a sharp transition to a non-topological phase when $\Delta_{\rm eff} \tau_{\rm eff} = 1/2$. At the critical disorder strength the density of states $\nu(\varepsilon)$ is singular \cite{dyson53,motrunich01,brouwer00,gruzberg05}
\begin{equation}
  \nu(\varepsilon) \propto |(v_{\rm F} \varepsilon/\Delta_{\rm eff}) \ln^3(\varepsilon/\Delta_{\rm eff})|^{-1},
  \label{eq:dyson}
\end{equation}
whereas for weaker disorder $\nu(\varepsilon)$ has a power-law tail with an exponent that depends on $\tau_{\rm eff}$ \cite{motrunich01,gruzberg05},
\begin{equation}
  \nu(\varepsilon) \propto \varepsilon^{4 \Delta_{\rm eff} \tau_{\rm eff} - 3 }.
\end{equation}
Although both the topological and the non-topological phases have a gapless excitation spectrum in the presence of disorder, the topological phase distinguishes itself by the presence of Majorana end states at an energy that is exponentially small in the wire length $L$.  \cite{motrunich01} Alternatively, for a wire that is coupled to normal-metal leads, the topological phase is characaterized by a negative sign of the determinant of the reflection matrix $r$, whereas $\det r$ is positive for the nontopological phase. \cite{akhmerov11} 

These results allow us to derive the phase diagram of the quantum wire as function of Zeeman coupling and disorder in some detail.
Consider first the regime discussed in Section\ \ref{(iii)} of large Zeeman fields at fixed $\mu$, for which $1/\tau_{\rm eff} = \gamma (m/2B)^{1/2}$ and $\Delta_{\rm eff} = 2 \Delta (\varepsilon_{\rm so}/B)^{1/2}$. Hence, the phase boundary between the topological and non-topological phases is at the critical disorder strength $\gamma_{\rm cr} = 4 \Delta u$. The critical disorder strength decreases with the Zeeman field due to the concurring suppression of the $p$-wave gap $\Delta_{\rm eff}$ when considering the case of fixed density, where $1/\tau_{\rm eff} = \gamma/u$ and $\Delta_{\rm eff} = 4 \Delta (\varepsilon_{\rm so}/B)$ so that $\gamma_{\rm cr} = 8 \Delta u (\varepsilon_{\rm so}/B)$.

In the intermediate regime presented in Section\ \ref{(ii)}, the contribution to $V_{\rm eff}(x)$ which is linear in $V(x)$ gives rise to an elastic scattering rate $1/\tau_{\rm eff} = \gamma (B/4 \varepsilon_{\rm so})^2/u$. The associated critical disorder strength is $\gamma_{\rm cr} = 2 u \Delta (4 \varepsilon_{\rm so}/B)^2$. This result is appropriate as long as $\gamma_{\rm cr} \ll \gamma_{\rm lin}$, which holds for $B\gg(\Delta/\varepsilon_{\rm so})^{1/5} \varepsilon_{\rm so}$. The expression for $\gamma_{\rm cr}$ decreases with the Zeeman field, so that the critical disorder strength does not increase  with $B$ throughout the entire range of $B$ where the phase boundary between topological and non-topological phases is governed by the disorder contribution linear in the bare disorder potential. In contrast, the critical disorder strength increases with $B$ when $B \ll (\Delta/\varepsilon_{\rm so})^{1/5} \varepsilon_{\rm so}$, where the quadratic contribution to $V_{\rm eff}$ is dominant. The quadratic contribution to $V_{\rm eff}$ gives an elastic scattering rate of $1/\tau_{\rm eff} = \gamma^2/2Bu^2$. This yields $\gamma_{\rm cr} = 2(\Delta B u^2)^{1/2}$ for the critical disorder strength.

In the regime of weak Zeeman fields (c.f.\ Section\ \ref{(i)}), the low-energy model in Eq.\ (\ref{WeakZeeman}) maps onto a random-hopping chain with staggered mean hopping. \cite{dyson53} In this model, an infinite disorder strength is needed to tune to the critical point if a finite gap is present in the non-disordered chain, \cite{ovchinnikov77,bouchaud90} indicating that any finite disorder strength preserves the topological phase. However, we still find a dependence of the phase boundary on disorder due to a more basic effect: Since the effective disorder potential is quadratic in the bare disorder $V(x)$, the disorder-induced Zeeman field $w(x)$ has not only a random component, but also a finite average $\langle w(x)\rangle = \gamma/u$. Due to this average, the phase boundary is no longer given by $b=0$, but rather determined by $b-\gamma/u=0$, or $B=\Delta + \gamma/u$. The density of states shows a power-law dependence on both sides of the topological phase transition,  with
\begin{equation}
  \nu(\varepsilon) \propto \varepsilon^{\alpha-1},\ \
  \mbox{with}\ \alpha = \gamma^{-2} |4 u \Delta ( b u - \gamma)|.
\end{equation}
Exactly at the phase transition, $\nu(\varepsilon)$ has the Dyson singularity (\ref{eq:dyson}), \cite{dyson53,ovchinnikov77,bouchaud90} with $\Delta_{\rm eff} = \gamma^2/4 u^2 \Delta$. Of course, these conclusions are restricted to disorder strengths smaller than the maximal disorder strength $\gamma_{\rm max} \sim u \Delta$, where the mapping to the Hamiltonian in Eq.\ (\ref{WeakZeeman}) is valid. In fact, the matching with the regime of intermediate Zeeman fields shows that the topological phase is lost once disorder becomes of the order of $\gamma_{\rm max}$.

We have also computed the  critical disorder strength numerically with a method based on a scattering approach to the full model~(\ref{BdG}).
The method, that is explained in detail in Appendix~\ref{numerics}, consists in computing numerically the reflection matrix $r(L)$ of a wire of length $L$.
 The existence of the a topological phase (i.e. the presence of Majorana bound states) is then signaled by the condition $\det[r(L)]=-1$ [as opposed to $\det [r(L)]=1$ for a non-topological gapped phase].  \cite{Fulga2011}

\begin{figure}[t]
	\begin{minipage}{0.45\linewidth} \includegraphics[width=0.9\linewidth, keepaspectratio=true]{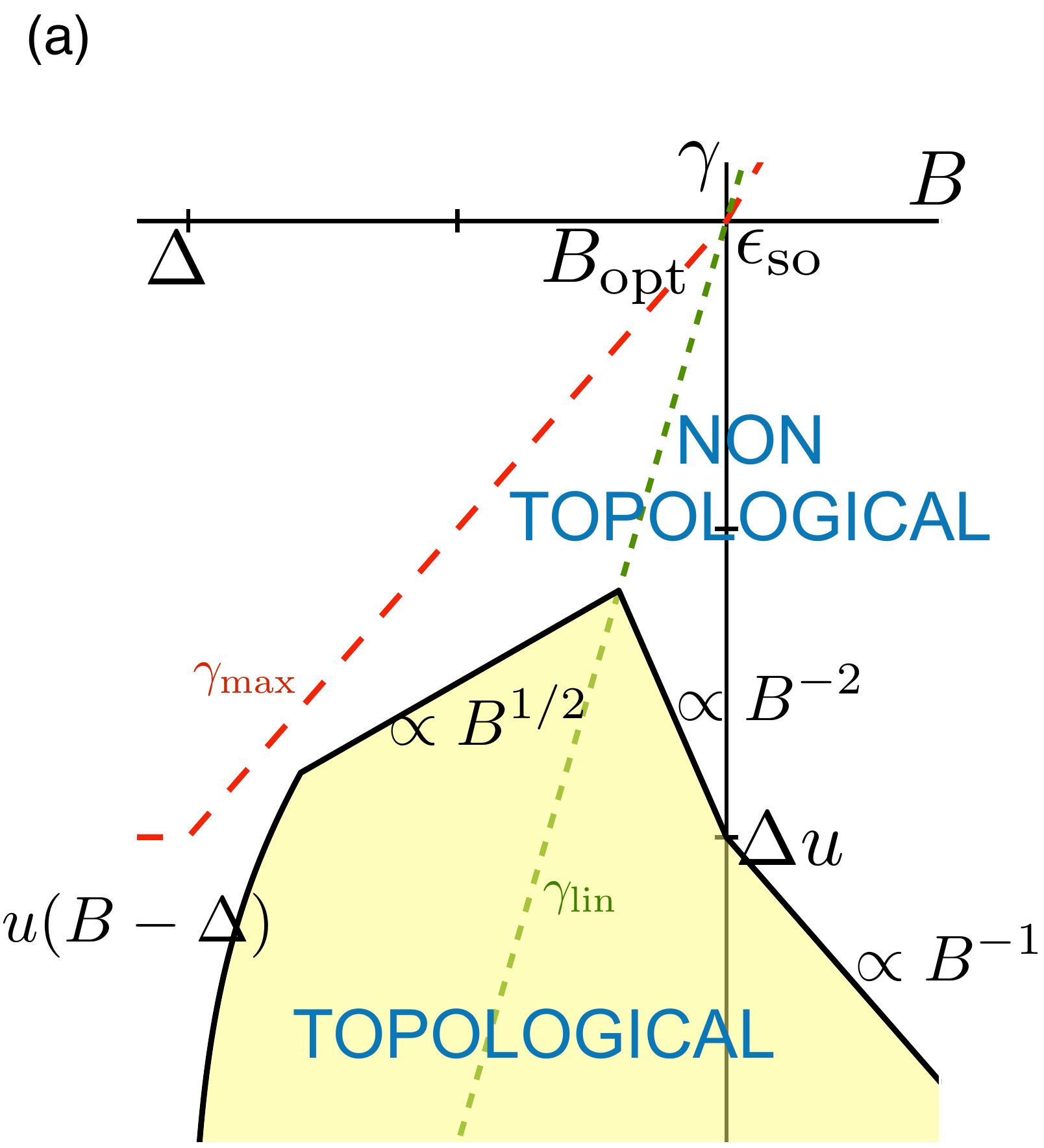}  \end{minipage} 	\begin{minipage}{0.45\linewidth} \includegraphics[width=1.0\linewidth, keepaspectratio=true]{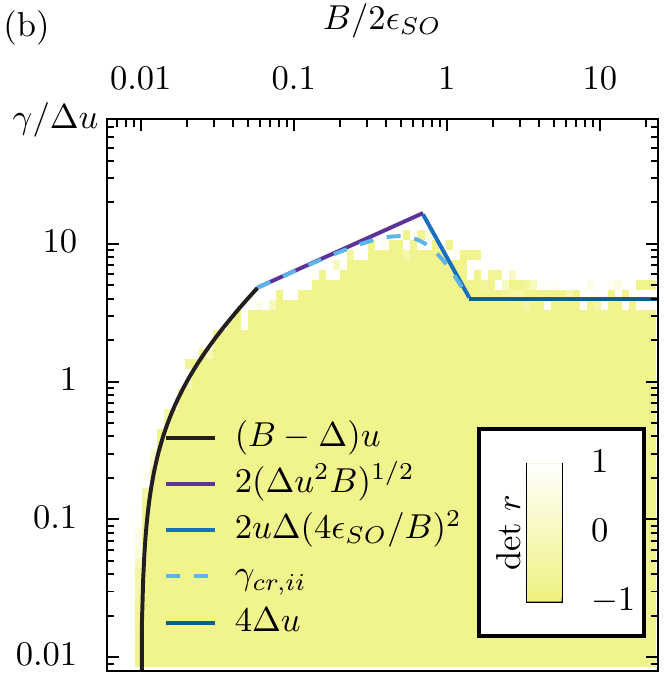}  \end{minipage}
\caption{(Color online)
Phase diagrams of the quantum wire as function of Zeeman field $B$ and disorder strength $\gamma$ in a log-log plot (a) for fixed density $n$ and (b) for fixed chemical potential including numerical results. In (a), the long-dashed line delineates the limit of validity of the low-energy models; at the short-dashed line, $V_{\rm eff}(x)$ changes between linear and quadratic dependence on the bare disorder $V(x)$. In (b), the numerics is based on calculating the determinant of the reflection matrix for the full Hamiltonian (\ref{BdG}), cp.\ Ref.\ \onlinecite{akhmerov11}. $\gamma_{cr,ii}$ (dashed line) interpolates between the linear and quadratic contributions to $\gamma_{\rm eff}$ for intermediate Zeeman fields.
\label{fig:phasediagram}}
\end{figure}

\section{Conclusion}

Our results for the critical disorder strength vs Zeeman field are summarized in Fig.\ 2 showing both the analytical and numerical results for the phase boundary.
This phase diagram emphasizes the important implication that the topological superconducting phase is most stable against disorder at the optimal Zeeman field $B_{\rm opt}\sim(\Delta/\varepsilon_{\rm so})^{1/5} \varepsilon_{\rm so}$. This is valid for fixed chemical potential [Fig.\ \ref{fig:phasediagram}(b)], but most pronounced for fixed density [Fig.\ \ref{fig:phasediagram}(a)], where the critical disorder strength $\gamma_{\rm cr} \to 0$ as $B\to \infty$.

As discussed above, the cause of this effect is the vanishing of the effective $p$-wave gap $\Delta_{\rm eff}$ in the limit $B \to \infty$. It is instructive to compare the large-$B$ limit of the semiconductor model considered here with that of a half-metallic ferromagnet, brought into electrical contact with a superconductor with spin-orbit coupling. \cite{duckheim11,chung11} In both models, charge carriers are fully spin polarized. Yet, the induced gap $\Delta_{\rm eff}$ is independent of the exchange field for the half metals of Refs.\ \onlinecite{duckheim11,chung11}, the reason being that $\Delta_{\rm eff}$ is set by the strength of the spin-orbit coupling in the superconductor. Since $\Delta_{\rm eff}$ directly determines the critical disorder strength $\gamma_{\rm cr}$, we are thus lead to expect that even at fixed density, the critical disorder strength $\gamma_{\rm cr}$ for a semiconductor wire can be made to saturate in the limit $B \to \infty$, if the helical state in the semiconductor wire is attributed to spin-orbit coupling in the superconductor, rather than spin-orbit coupling intrinsic to the semiconductor.

\acknowledgments

We acknowledge discussions with Yuval Oreg and financial support by the Deutsche Forschungsgemeinschaft through SPP 1285 as well as by the Alexander von Humboldt foundation. Upon completion of this manuscript Ref.\ \onlinecite{potter11} appeared, which discusses the influence of impurities in the superconductor on the proximity effect. 

\appendix

\section{Numerical analysis of the topological phase}
\label{numerics}

In this appendix, we describe the algorithm that allows us to find the
boundary between the normal and the topological phase in a
superconducting quantum wire with spin-orbit interaction. The
topological phase is marked by the existence of bound Majorana end
states. Knowledge of the reflection matrix $r(L)$ of the wire (with
length $L$) allows one to determine whether Majorana states exist and,
thus, to determine the topological properties of the
wire. Specifically, an odd number of Majorana bound state exists
precisely if $\det r = -1$. \cite{Fulga2011} The numerical analysis
thus focuses on the scattering matrix $S(L)$ of the wire and its reflections
submatrix $r(L)$.

\begin{figure}[t]
  \centering
    \includegraphics[width = 0.9 \linewidth]{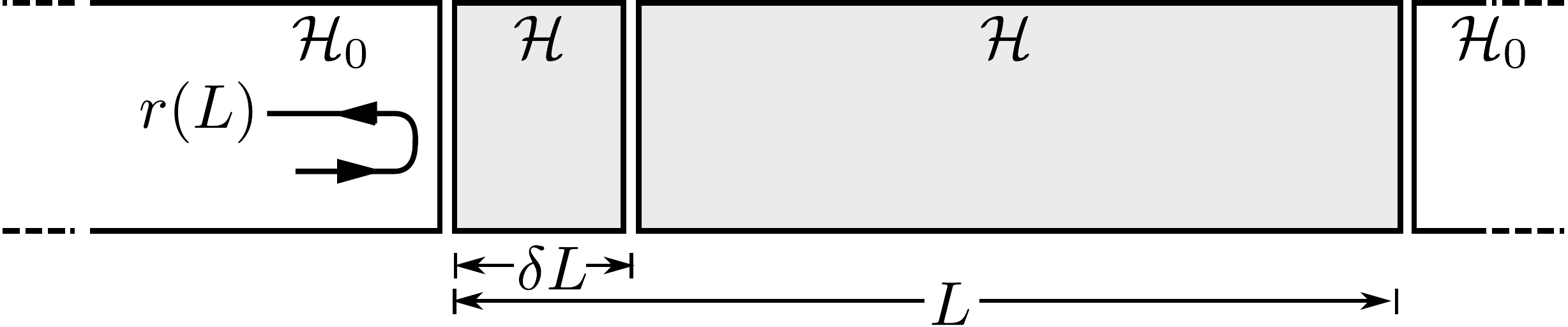}
    \caption{Quantum wire of length $L$ (grey shading) described by
      Hamiltonian ${\cal H} $, Eq.~(\ref{eq:BdG-wire}), with
      unperturbed leads(white) described by the Hamiltonian ${\cal
        H}_0$. The reflection matrix $r(L)$ is found by repeatedly
      concatenating scattering matrices of segments with length
      $\delta L$.}
  \label{fig:setup}
\end{figure}

The situation we consider is shown in Fig.~\ref{fig:setup}.  The
scattering states are chosen as the eigenstates of the Hamiltonian
${\cal H}_0$ of a 'free' wire (lead) with $u=0$, $\Delta=0$, $B=0$,
but with finite chemical potential $\mu_0$,
\begin{align}
  {\cal H}_0 & = \left( \frac{p^2}{2m}  - \mu_0 \right)\tau_z 
  \label{BdG-lead} \, .
\end{align}

The lead is attached to the central wire segment with nonzero $u$,
$\Delta$, $B$, and chemical potential $\mu$. The Hamiltonian of the
central segment is given by
\begin{align}
  \label{eq:BdG-wire}
  {\cal H} & =   {\cal H}_0   +   \delta {\cal H}  \, ,
\end{align}
with
\begin{align}
 \delta {\cal H} & = \Delta \tau_x + V(x) \tau_z + u p \sigma_x \tau_z
 - B  \sigma_z  + \left(  \mu_0 -\mu \right)\tau_z   \, .
\end{align}
Here the symbols $u$, $V(x)$, $\Delta$, and $B$ are explained in the
main text. The scattering matrix of a segment of length $L$ of the
wire has the form
\begin{align}
  S(L) = \left(
    \begin{array}{cc}
      t(L) & r'(L)\\
      r(L) & t'(L)
    \end{array}
\right) \, ,
\end{align}
where $t$, $r$ and $t'$, $r'$ are transmission and reflection matrices
for scattering states incoming from the left and the from the right,
respectively. The scattering matrix of two wire segments attached to
each other, whose individual scattering matrices are $S_1$ and $S_2$,
is given by
\begin{align}
\label{eq:s-matrix-multiplication}
  S_{12}  & = S_1 \otimes S_2 = \left(
    \begin{array}{cc}
      t_{12} & r'_{12}\\
      r_{12} & t'_{12}
    \end{array}
\right)  \, ,
\end{align}
where
\begin{align}
t_{12} & = t_2 [{\bf 1} - r'_1 r_2]^{-1} t_1 \, ,\\
r_{12} & = r_1 + t'_1 [{\bf 1} - r_2 r'_1]^{-1} r_2 t_1 \, ,\\
t'_{12} & = t'_1 [{\bf 1} - r_2 r'_1]^{-1} t'_2 \, ,\\
r'_{12} & = r'_2 + t_2 [{\bf 1} - r'_1 r_2]^{-1} r'_1 t'_2 .
\end{align}


The algorithm to find $S(L)$ consists of repeated concatenations of
scattering matrices. The wire with length $L$ is split into segments
of length $\delta L$ and the scattering matrices $ S_\delta = S(\delta
L)$ of the segments are combined according to
Eq.~(\ref{eq:s-matrix-multiplication}). This method has been used
previously to study disordered wires with unconventional
superconductivity \cite{Brouwer2003} and conductivity scaling in
graphene. \cite{Bardarson2007} In the limit where the segment length
$\delta L$ is small (such that $W \equiv S_\delta - {\bf 1} $ deviates
only slightly from the zero matrix) the scattering matrix of a segment
can be calculated analytically in linear order in $\delta L$ (see
subsection below). [Note that this approximation is not based on the
smallness of the perturbations $ \delta {\cal H}$ but on the smallness
of $\delta L$.]

The initial condition $S(0)$ is to taken to be  (open wire)
    \begin{align}
      S &= \left(
        \begin{array}{cc}
          {\bf 1} & {\bf 0}\\
          {\bf 0} & {\bf 1}
        \end{array}
      \right)     \, .
\end{align} 
In order to preserve unitarity to all orders, we set
\begin{align}
  S_\delta  \to S_\delta  = \left({\bf 1} + \frac{1}{2} W \right) \left({\bf 1}
    - \frac{1}{2} W \right)^{-1} \label{eq:cayley-trans} \, ,
\end{align}
where $W = S_\delta - {\bf 1}$ contains the leading-in-$\delta L$
contributions.

\subsection{Scattering matrix of a small segment}

To linear order in the segment length $\delta L$ the scattering matrix $
S_\delta$ can be obtained analytically. The scattering states in the leads
are indexed by electron-hole character $\sigma=\pm1$, propagation
direction (right-moving, left-moving) $\rho=\pm1$, spin in z-direction
(up, down) $s=\pm1$, and energy measured from the Fermi energy,
$\epsilon$. They are of the form
\begin{align}
  \psi_{\sigma,\rho,s,\epsilon}=\frac{e^{i \sigma \rho
      k_{\sigma, \epsilon}x}}{\sqrt{v_{\sigma, \epsilon}}}\chi_{\sigma, s}
\end{align}
where $k_{\sigma,\epsilon}=\sqrt{2m(\mu_0 + \sigma \epsilon)}$ is the
momentum, $v_{\sigma, \epsilon} = k_{\sigma,\epsilon} / m$ the
velocity, and $\chi $ is a 4-component spinor in Nambu- and spin
space. To linear order in $\delta L$ the contribution $ S_\delta$ from
each term $\delta {\cal H}$ can be calculated individually. The
contributions from the deterministic terms in $\delta {\cal }H$ are
obtained by matching the scattering states
$\psi_{\sigma,\rho,s,\epsilon}$ at the interface between the leads and
the wire to eigenfunctions of the wire segment.  For the disorder
potential $V(x)$ the contribution to $S_\delta $ is random and, to
leading order in $\delta L$, can be obtained using the first-order
Born approximation
\begin{align}
  \delta r_{\sigma \sigma',  s s'}^{V}  & = -\frac{i}{\hbar} \langle
  \psi_{\sigma',-,s',\epsilon}| V \tau_z| \psi_{\sigma,+,s,\epsilon}
  \rangle  \\  & = \frac{-i}{\hbar} \int_{0}^{\delta L}   \frac{ V(x) e^{i \left(\sigma' k_{\sigma', \epsilon} + \sigma 
        k_{\sigma, \epsilon} \right)x} \delta_{s,s'}
    \delta_{\sigma,\sigma'}}{\sqrt{v_{\sigma, \epsilon} v_{\sigma',
        \epsilon}}}   dx \notag \\
  \delta t_{\sigma \sigma',  s s'}^{V}  & = -\frac{i}{\hbar} \langle
  \psi_{\sigma',+,s',\epsilon}| V \tau_z| \psi_{\sigma,+,s,\epsilon}
  \rangle  \\  & = \frac{-i}{\hbar} \int_{0}^{\delta L}   \frac{ V(x) e^{i \left(\sigma' k_{\sigma', \epsilon} - \sigma 
        k_{\sigma, \epsilon} \right)x} \delta_{s,s'} \delta_{\sigma,\sigma'}}{\sqrt{v_{\sigma, \epsilon} v_{\sigma', \epsilon}}}   dx \notag
\end{align}
and similarly for $t'$ and $r'$. From the correlator $\langle
V(x)V(x') \rangle = \gamma \delta(x -x')$ one finds
\begin{align}
 \langle | \delta r_{\sigma \sigma',  s s'}^{V}|^2 \rangle  = \langle  | \delta t_{\sigma
   \sigma',  s s'}^{V}|^2 \rangle =  \delta_{s,s'}
 \delta_{\sigma,\sigma'} \frac{\delta L}{(\hbar v_F)^2} \gamma ,
 \label{eq:statistics}
\end{align}
where $v_F = v_{\sigma, \epsilon=0}$ is the Fermi velocity. For each
segment the contribution of $V$ to $S_\delta $ is thus taken as a
random matrix with zero average and mean square given by
Eq.~(\ref{eq:statistics}).

\bigskip

\subsection{Phase diagram}

Since the wire has a superconducting gap the reflection part, $r(L)$,
of the scattering matrix becomes unitary in the limit of a long wire
and for $\epsilon = 0$ the determinant of $r(L)$ approaches either
$\det r \to + 1$ or $\det r \to - 1$ with the latter indicating the
existence of a Majorana state. To obtain the phase diagram the
algorithm is repeated for different magnetic fields $B$ and different
disorder strengths $\gamma$ [but with a single disorder realization
for each disorder strength]. From this the determinant $\det r(L)$ is
calculated and plotted as a function of $B$ and $\gamma$.  We have
verified that the phase diagram is independent of the choice of the
chemical potential $\mu_0$ in the lead. The data is shown in Fig.~2b
together with the theoretically predicted phase boundaries.

\end{document}